# Effect of Electron-Electron Interactions on Metallic State in Quasicrystals

Shiro Sakai[1] and Akihisa Koga[2]

[1]*Center for Emergent Matter Science, RIKEN, Wako, Saitama 351-0198, Japan*
[2]*Department of Physics, Tokyo Institute of Technology, Meguro, Tokyo 152-8551, Japan*

**Abstract** We theoretically study the effect of electron-electron interactions on the metallic state of quasicrystals. To address the problem, we introduce the extended Hubbard model on the Ammann-Beenker tiling as a simple theoretical model. The model is numerically solved within an inhomogeneous mean-field theory. Because of the lack of periodicity, the metallic state is nonuniform in the electron density even in the noninteracting limit. We clarify how this charge distribution pattern changes with electron-electron interactions. We find that the intersite interactions change the distribution substantially and in an electron-hole asymmetric way. We clarify the origin of these changes through the analyses in the real and perpendicular spaces. Our results offer a fundamental basis to understand the electronic states in quasicrystalline metals.

## 1. Introduction

Quasicrystal is a solid with an aperiodic but regular arrangement of atoms.[1] The structure, characterized by a rotational symmetry and self-similarity, may give nontrivial consequences to its electronic state. In particular, the lack of periodicity leads to an inhomogeneous (but regular) electronic distribution reflecting the underlying quasiperiodic lattice structure. This inhomogeneous electronic state indeed shows peculiar properties like a critical wave function and singular-continuous spectrum,[2-9] as well as the multifractality,[10,11] already in the noninteracting limit.

The introduction of electron-electron interactions will produce further diverse and novel electronic states. For example, interesting magnetic structures have been found in the antiferromagnetic states in the Heisenberg[12-14] or Hubbard[15,16] models. Superconductivity with a nonuniform order parameter has also been found and discussed in the attractive[17-24] and repulsive[25] Hubbard models. Moreover, topological nature of the superconducting states has been studied.[25-27]

In this paper, we study the effect of electron-electron interactions on the *metallic* state of quasicrystals. In fact, most known quasicrystals are metallic alloys. Nevertheless, the effect of the electron-electron interactions on the metallic state has not been explored in detail, especially in two and three dimensions where the geometry of the lattice can be relevant. While the effect of onsite repulsion has been discussed in several previous works[28,29], here we are particularly interested in the effect of intersite interactions, which will directly depend on the geometry around each site. We will see that the intersite interactions produce charge distributions distinct from those present in the

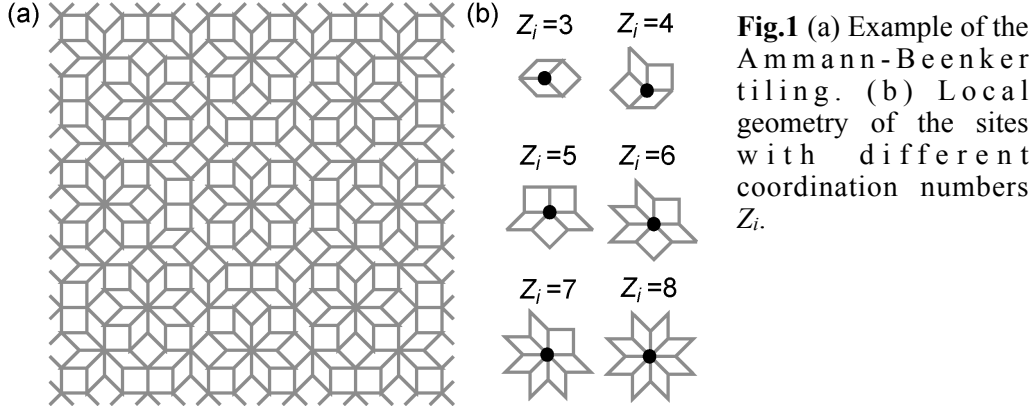

Fig.1 (a) Example of the Ammann-Beenker tiling. (b) Local geometry of the sites with different coordination numbers $Z_i$.

noninteracting system and that its effect works in an electron-hole asymmetric way. These clarifications would give a basis to understand the electronic states of quasicrystalline metals, including the recently discovered one in the twisted bilayer graphene.[30]

## 2. Model and Method
### 2.1 Extended Hubbard model on Ammann-Beenker tiling

The Ammann-Beenker tiling[31] is a prototype of two-dimensional quasiperiodic structures. The tiling consists only of squares and rhombuses, and possesses an eight-fold rotational symmetry [Fig. 1(a)]. We regard each vertex as a site. Then, different sites have different local geometries except when the rotational symmetry relates them. For instance, the coordination number $Z_i$ of a site $i$ ranges from 3 to 8, as illustrated in Fig. 1(b). We take the length of the edge of the squares and rhombuses as the length unit.

We study the extended Hubbard model on this lattice. The Hamiltonian reads

$$H = -t\sum_{\langle ij\rangle\sigma}\left(c^\dagger_{i\sigma}c_{j\sigma}+h.c.\right) - \mu\sum_{i\sigma}n_{i\sigma} + U\sum_i n_{i\uparrow}n_{i\downarrow} + V\sum_{\langle ij\rangle\sigma\sigma'}n_{i\sigma}n_{j\sigma'}, \quad (1)$$

where $c^\dagger_{i\sigma}$ ($c_{i\sigma}$) creates (annihilates) an electron of spin $\sigma$ at site $i$ and $n_{i\sigma}\equiv c^\dagger_{i\sigma}c_{i\sigma}$. The first term represents the electron hopping between the neighboring sites (denoted by $\langle ij\rangle$) connected by the edge of the square or the rhombus. We take $t=1$ as the unit of energy. For this choice, the bare "bandwidth" of the site-averaged density of states is about $8.5t$. The third and fourth terms represent the onsite and nearest-neighbor Coulomb interactions, respectively. The chemical potential $\mu$ is determined self-consistently to fix the average electron density, $\bar{n}\equiv\frac{1}{N}\sum_i n_i$ with $n_i\equiv\sum_\sigma\langle n_{i\sigma}\rangle$, at a given value. Here, $\langle\ldots\rangle$ denotes the expectation value in the ground state. Note that this tiling is bipartite so that there is an electron-hole symmetry (i.e., the system of $\bar{n}$ is equivalent to that of $2-\bar{n}$ through the electron-hole transformation) unless $V$ is finite.

### 2.2 Hartree-Fock approximation and its implementation

Within the Hartree-Fock approximation, the interaction part of the Hamiltonian (1) is reduced to

$$\sum_{i\sigma}\left[U\langle n_{i\bar{\sigma}}\rangle + V\sum_{j:\text{n.n. of }i} n_j\right] n_{i\sigma} - V\sum_{\langle ij\rangle\sigma}\langle c_{i\sigma}^\dagger c_{j\sigma}\rangle c_{j\sigma}^\dagger c_{i\sigma}. \qquad (2)$$

The first and second terms represent the Hartree and Fock contributions, respectively. While the $U$ term gives a site-dependent potential only through the nonuniform distribution of $n_i$, the $V$ term of the Hartree type gives a more strongly site-dependent potential because it directly depends on the connectivity of the sites. The $V$ term of the Fock type modulates the electron-hopping amplitude in a site-dependent way.

To solve the mean-field Hamiltonian for a large-size cluster, we employ the kernel polynomial method,[32] which expands the local density of states in the Chebyshev polynomials. The algorithm is further accelerated by utilizing the idea of the localized Krylov subspace,[23] which reduces the dimension of the Hilbert space to be taken into account in the computation of the coefficients in the Chebyshev expansion. From the obtained local density of states, we calculate the electron density $n_i$, which updates the mean-field Hamiltonian. We iterate the procedure until the convergence is obtained. The calculations are done for 134241-site and 781057-site clusters at zero temperature.

## 2.3 Perpendicular space

In order to see the effect of the geometry around each site systematically, it is convenient to plot the local quantities (like $n_i$) in the perpendicular space, which consists of the dimensions perpendicular to the physical space in a high-dimensional description of the quasiperiodic lattice: When the Ammann-Beenker tiling is obtained through projecting a four-dimensional hypercubic lattice onto a two-dimensional plane, the perpendicular space is defined by the remaining two dimensions.

Suppose the lattice points on the four-dimensional hypercubic lattice is represented by $\vec{m} = \sum_{i=1,2,3,4} m_i \vec{d}_i$ with integers $\{m_i\}$ and four-dimensional lattice vectors $\{\vec{d}_i\}$. To obtain the Ammann-Beenker tiling, we project the vector $\vec{m}$ onto the physical space spanned by $\vec{e}_x$ and $\vec{e}_y$ with $(\vec{e}_x)_j = \cos(j\pi/4)$ and $(\vec{e}_y)_j = \sin(j\pi/4)$ for $j = 1,2,3,4$. Namely, the two-dimensional vectors $\vec{r} = (x,y) = (\vec{m}\cdot\vec{e}_x, \vec{m}\cdot\vec{e}_y)$ constitute the Ammann-Beenker tiling. On the other hand, the perpendicular space is spanned by the four-dimensional vectors $\vec{e}_{\tilde{x}}$ and $\vec{e}_{\tilde{y}}$ with $(\vec{e}_{\tilde{x}})_j = \cos(3j\pi/4)$ and $(\vec{e}_{\tilde{y}})_j = \sin(3j\pi/4)$ for $j = 1,2,3,4$. Namely, the coordinate in the perpendicular space is given by $\vec{\tilde{r}} = (\tilde{x}, \tilde{y}) = (\vec{m}\cdot\vec{e}_{\tilde{x}}, \vec{m}\cdot\vec{e}_{\tilde{y}})$.

Because the sites in a similar geometry in the physical space are located in a similar geometry in the high-dimensional space, too, they possess a similar coordinate in the perpendicular space. In fact, the perpendicular space is divided into sections, each of which corresponds to a particular local geometry in the physical space. Plotting in the perpendicular space is therefore convenient to discuss the dependence on the local geometry. We utilize this plot in Sec.3.4.

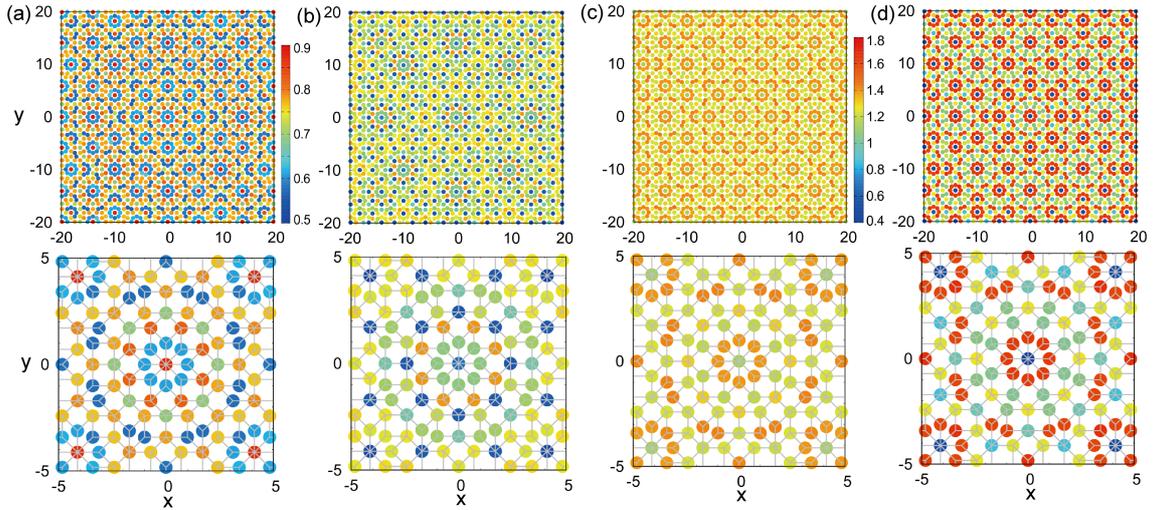

**Fig.2** Real-space map of the electron density for (a) $\bar{n} = 0.7$ and $U=V=0$, (b) $\bar{n} = 0.7$ and $U=4V=2$, (c) $\bar{n} = 1.3$ and $U=V=0$, and (d) $\bar{n} = 1.3$ and $U=4V=2$. Panels (a) and (b) [(c) and (d)] share the color scale. The calculations were done for a cluster with $N$=134241 sites, among which the central area satisfying |x|, |y|<20 (5) is plotted in top (bottom) panels.

## 3. Results and Discussions
### 3.1 Real-space map of electron-density distribution

First, we show that the electron-density distribution changes its spatial pattern, according to the average filling and interaction strength. Figure 2 plots $n_i$ in real space for various parameter sets. Already for $U=V=0$ [panels (a) and (c)], the electron density shows nonuniform distributions with the eight-fold rotational symmetry, reflecting the underlying lattice structure. However, the introduction of the interactions, especially the effect of $V$, alters the patterns substantially.

As we can see in the lower panels, from panels (a) to (b) (for $\bar{n} = 0.7$), the red (high-density) sites with the coordination number $Z_i$=8 change to blue (low density) while blue or light-blue sites with $Z_i$=3 gain population to become yellow - light green (medium density). On the other hand, from panels (c) to (d) (for $\bar{n} = 1.3$), the interaction effect appears to enhance the contrast: The light-green (slightly low density) sites with $Z_i$=8 become blue (very low density) while the orange (slightly high density) sites with $Z_i$=3 become red (high density).

Figure 3 shows the distribution of $n_i$ for the corresponding parameter sets. For $U=V=0$ [panels (a) and (c), which are equivalent under the electron-hole transformation], there are several groups in the range from 0.5 to 0.9. As the interaction is introduced, more sites are accumulated to $n_i$=0.7-0.8 for $\bar{n} = 0.7$ while the distribution spreads more for $\bar{n} = 1.3$.

Thus, the electron-electron interactions have a significant effect on the electron distributions on the quasiperiodic lattice. The effect can be even more significant and nontrivial than that in periodic systems since it acts differently on different sites in quasiperiodic systems. Moreover, the above results manifest different interaction effects

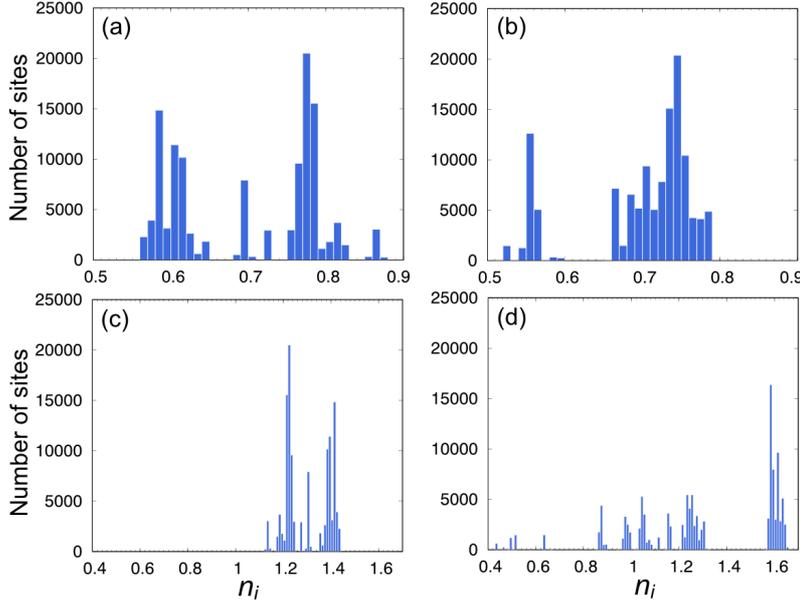

**Fig.3** Distribution of the electron density for (a) $\bar{n} = 0.7$ and $U=V=0$, (b) $\bar{n} = 0.7$ and $U=4V=2$, (c) $\bar{n} = 1.3$ and $U=V=0$, and (d) $\bar{n} = 1.3$ and $U=4V=2$. The calculations were done for 134241 sites, among which we have used only the sites with $\sqrt{x^2 + y^2} < 180$ to avoid the boundary effect. The bin width is set to be 0.01.

between the $\bar{n} < 1$ and $\bar{n} > 1$ cases, breaking the electron-hole symmetry.

### 3.2 Effect of local geometry

As the above results suggest, the coordination number $Z_i$ plays an important role in determining the charge distribution patterns. This is also expected from Eq. (2), where the $V$ term explicitly depends on the local geometry around each site.

To clarify this point, we plot in Fig.4 the electron density $n_i$ against $Z_i$ of each site $i$. For $\bar{n} = 0.7$ [panel (a)], we find a tendency that the sites with a larger $Z_i$ have a larger electron density for $U=V=0$ (black squares). The effect of $U$ mitigates this tendency (red circles) because it prefers a uniform distribution to reduce the onsite-interaction energy. On the other hand, the effect of $V$ completely changes the tendency (blue triangles): Now, the sites with a larger $Z_i$ tend to have a smaller electron density. Physically, this is because the sites with a larger $Z_i$ tend to have a larger energy increase due to the $V$ term and thereby lose the population.

This effect of $V$ can be seen for $\bar{n} = 1.3$ [panel (b)], too. In this case, however, already for $V=0$, there is a tendency for the sites with a larger $Z_i$ to have a smaller electron density. This is compatible with the effect of $V$, explaining the enhancement of the contrast from Fig. 2(c) to 2(d), discussed in the previous section.

Thus, we have seen that the nonuniformity for $V=0$ mainly originates from the effect of $t$, where the opposite tendencies between $\bar{n} = 0.7$ and $\bar{n} = 1.3$ can be understood by the electron-hole symmetry around $\bar{n} = 1$. The effect of $V$, however, always prefers a less population at a site with a larger $Z_i$, breaking this electron-hole symmetry. This explains why the charge modulation for $\bar{n} = 1.3$ shows a much stronger amplitude than that for $\bar{n} = 0.7$: For the latter, the two tendencies (that due to $t$ and that due to $V$) compete while for the former they cooperate.

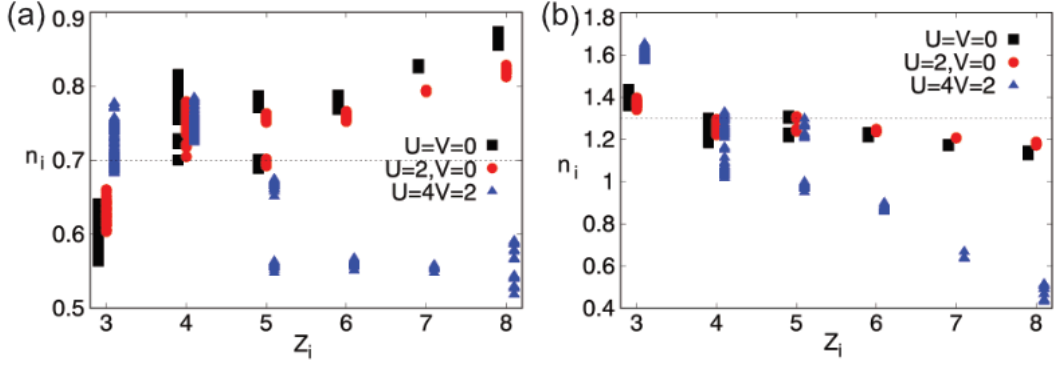

**Fig.4** Site occupation plotted against the coordination number $Z_i$. (a) $\bar{n} = 0.7$ and (b) $\bar{n} = 1.3$. Dotted horizontal lines show the value of $\bar{n}$. The calculations were done for 134241 sites, among which we have used only the sites with $\sqrt{x^2 + y^2} < 180$ to avoid the boundary effect.

### 3.3 Origin of electron-hole asymmetry

Because the Ammann-Beenker tiling is bipartite, there is an electron-hole symmetry for $V$=0: The behavior of holes for $\bar{n} = 1 - \delta$ is equivalent to that of electrons for $\bar{n} = 1 + \delta$. This is understood by the electron-hole transformation, $c_{i\sigma}^\dagger \to (-1)^i h_{i\sigma}$ and $c_{i\sigma} \to (-1)^i h_{i\sigma}^\dagger$, where $h_{i\sigma}^\dagger (h_{i\sigma})$ denotes the hole creation (annihilation) operator and the factor $(-1)^i$ changes the sign when the sublattice changes. Namely, this transformation does not change the shape of the Hamiltonian except for the replacement of $c$ with $h$ and the shift of the chemical potential. While in the periodic systems, this electron-hole symmetry holds even under a finite $V$, it is broken in the quasiperiodic systems, as we have seen that $V$ gives $n_i$ a tendency decreasing with $Z_i$ for both $\bar{n} < 1$ and $\bar{n} > 1$.

To see the origin of this electron-hole asymmetry, we consider the electron-hole transformation for the intersite interaction in Eq.(1), which is transformed as

$$V \sum_{\langle ij \rangle \sigma\sigma'} n_{i\sigma} n_{j\sigma'} \to -2V \sum_{i\sigma} Z_i n_{i\sigma}^h + V \sum_{\langle ij \rangle \sigma\sigma'} n_{i\sigma}^h n_{j\sigma'}^h, \qquad (3)$$

with $n_{i\sigma}^h \equiv h_{i\sigma}^\dagger h_{i\sigma}$. The first term on the right-hand side gives a site-dependent potential, which cannot be absorbed into the chemical potential, in contrast to the periodic systems. This is the source of the electron-hole asymmetry, always giving a lower electron density for a site with a larger $Z_i$.

### 3.4 Perpendicular space

To elucidate the relation between the charge density and the geometry around a site, we plot $n_i$ in the perpendicular space. As is elaborated in Sec. 2.3, the sites in a similar geometry are assembled into the same area in the perpendicular space.

Figure 5 shows the map of $n_i$ in the perpendicular space. Each section separated by solid black lines has a specific coordination number $Z_i$, as denoted in the top panel of Fig. 5(a). The corresponding local geometries are illustrated in Fig. 1(b). While we can see that $Z_i$ certainly plays an important role in determining $n_i$, the color varies in each section, too, reflecting the variation of the geometry beyond nearest neighbors.

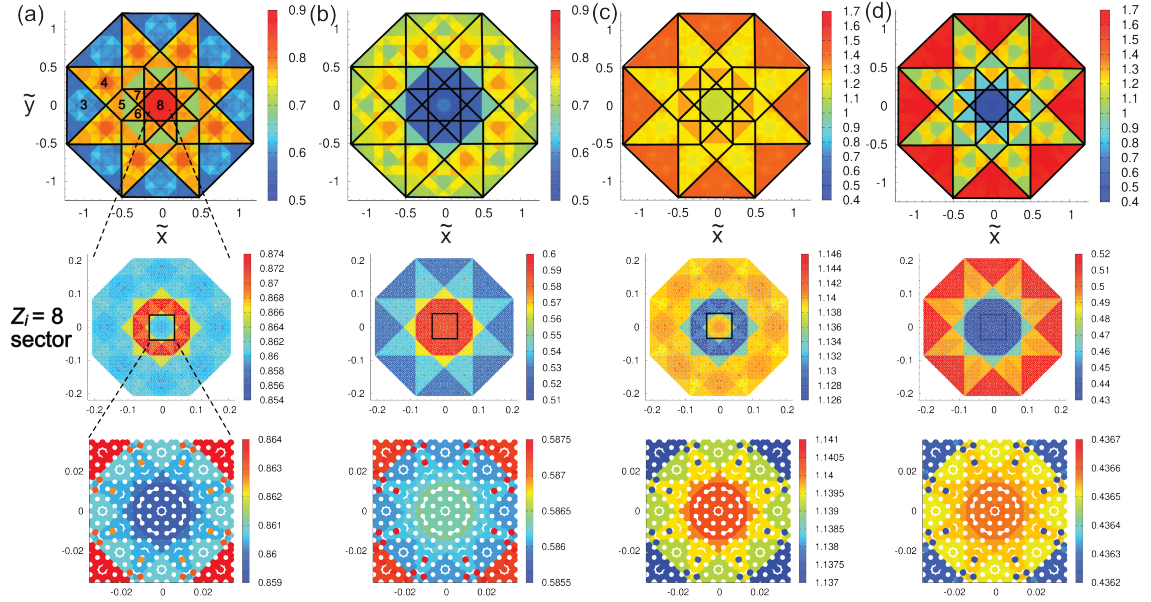

**Fig.5** Intensity map of the electron density in the perpendicular space for (a) $\bar{n} = 0.7$ and $U=V=0$, (b) $\bar{n} = 0.7$ and $U=4V=2$, (c) $\bar{n} = 1.3$ and $U=V=0$, and (d) $\bar{n} = 1.3$ and $U=4V=2$. Top panels show the map of the full perpendicular space, where the black lines separate it into areas specified by the coordination number [denoted in panel (a)] of the corresponding sites in the physical space. See Fig.1(b) for the corresponding local geometries. The middle panels show the enlarged maps of the $Z_i$=8 sector. The bottom panels show further enlarged maps around the center [denoted by a black square in the middle panels]. The calculations were done for 781057 sites, among which we used only the sites satisfying $\sqrt{x^2 + y^2} < 440$ to avoid the effect of boundary.

For $\bar{n} = 0.7$ [panels (a) and (b)], the charge distribution pattern is totally changed by the interaction. While $n_i$ is distinctive between the sections of $Z_i$=3 and 4 for $U=V=0$, it becomes less distinctive for $U=4V=2$. A similar change can be seen for the sections of $Z_i$=6, 7, 8 and a half section of $Z_i$=5. The electron-electron interactions increase the electron density in the $Z_i$=3 sections and decrease it in the other sections. For $\bar{n} = 1.3$ [panels (c) and (d)], the flow of the electron density is even more significant, as is evident in the larger scales of the color bar. Note that the patterns (a) and (c) (for $U=V=0$) are connected by the electron-hole transformation although they may look different due to the different color scales used here. With the interaction, the sections of $Z_i$=6, 7, 8 lose the electron density significantly while the $Z_i$=3 section gains popularity. The variation is larger for $\bar{n} = 1.3$ than for $\bar{n} = 0.7$ because the effects of $t$ and $V$ cooperates in the former while they compete in the latter.

An interesting observation here is that the enlarged view of the $Z_i$=8 section (at the center of the top panels) shows a variation similar to that in the original full perpendicular space: In the middle panels, the $Z_i$=8 section shows an internal structure which looks divided into subsections like the original one although the scale of the variation is one order of magnitude smaller than the original one and the magnitude relation between the subsections does not necessarily correspond to the original one. Further enlarged views of the central part of the $Z_i$=8 section show a further finer structure (bottom panels), which also has a structure similar to the original one. These fine structures reflect the self-similarity of the Ammann-Beenker tiling. Namely, when we connect the sites with $Z_i$=8 in the physical space, it is again reduced to the Ammann-Beenker tiling consisting of larger tiles, as was pointed out in Ref.16. This "superlattice" will give a small variation of $n_i$ among the $Z_i$=8 sites. However, because the effective hopping and onsite/intersite interactions in this superlattice may have a different sign from the original one and be nonuniform in space, the modulation pattern will not be a simply rescaled one. It is remarkable that there remains a discernible fine structure despite this difference.

## 4. Conclusion

We have studied the extended Hubbard model on the Ammann-Beenker tiling as a model of the interacting electron system in quasicrystals. While the metallic state is always inhomogeneous due to the absence of the periodicity, the spatial distribution of the electron density changes substantially with the electron-electron interactions. We have found that, while the onsite repulsion suppresses the inhomogeneity, the intersite interaction $V$ completely changes the pattern or substantially enhances the inhomogeneity. The different effects of $V$ come from the competition or cooperation with the effect of the electron hopping $t$, depending on the dominance of holes or electrons in the system. We have also revealed that the charge distribution is largely determined by the local geometry (connectivity to the neighboring sites) of each site while the effects beyond it, in particular those of a self-similarity, have also been found through the analysis in the perpendicular space. Our results indicate interesting metallic states present in the interacting electron systems on quasicrystals and will serve as a basis to understand such states.

**Acknowledgments**
This work was supported by JSPS KAKENHI Grant No. JP16H06345, JP20H05279 (SS), JP19H05821, JP18K04678, and JP17K05536 (AK).